\documentclass[sigconf]{acmart}

\AtBeginDocument{%
  \providecommand\BibTeX{{%
    \normalfont B\kern-0.5em{\scshape i\kern-0.25em b}\kern-0.8em\TeX}}}

\copyrightyear{2021}
\acmYear{2021}
\setcopyright{acmcopyright}
\acmConference[WI-IAT '21]{IEEE/WIC/ACM International
Conference on Web Intelligence}{December 14--17, 2021}{ESSENDON, VIC, Australia}
\acmBooktitle{IEEE/WIC/ACM International Conference on Web Intelligence (WI-IAT
'21), December 14--17, 2021, ESSENDON, VIC, Australia}
\acmPrice{15.00}
\acmDOI{10.1145/3486622.3493985}
\acmISBN{978-1-4503-9115-3/21/12}

\usepackage{graphicx}

\acmSubmissionID{WI281}

\usepackage{color}

\begin{document}

\title[Product Information Browsing Support System]
{Product Information Browsing Support System Using Analytic Hierarchy Process}

\author{Weijian Li}
\affiliation{%
  \institution{Department of Computer Science, Graduate School of Engineering, Nagoya Institute of Technology}
  \streetaddress{Gokiso-cho, Showa-ku}
  \city{Nagoya}
  \state{Aichi}
  \country{Japan}
  \postcode{466-8555}
}
\email{lwj@toralab.org}

\author{Masato Kikuchi}
\affiliation{%
  \institution{Nagoya Institute of Technology}
  \streetaddress{Gokiso-cho, Showa-ku}
  \city{Nagoya}
  \state{Aichi}
  \country{Japan}
  \postcode{466-8555}
}
\email{kikuchi@nitech.ac.jp}

\author{Tadachika Ozono}
\affiliation{%
  \institution{Nagoya Institute of Technology}
  \streetaddress{Gokiso-cho, Showa-ku}
  \city{Nagoya}
  \state{Aichi}
  \country{Japan}
  \postcode{466-8555}
}
\email{ozono@nitech.ac.jp}

\renewcommand{\shortauthors}{Li, et al.}

\begin{abstract}
  Large-scale e-commerce sites can collect and analyze a large 
  number of user preferences and behaviors, and thus can recommend 
  highly trusted products to users. However, it is very difficult 
  for individuals or non-corporate groups to obtain large-scale 
  user data. Therefore, we consider whether knowledge of the decision-making domain can be used to obtain user preferences 
  and combine it with content-based filtering to design an information retrieval system. 
  This study describes the process of building a product information browsing support system with high satisfaction 
  based on product similarity and multiple other perspectives about products on the Internet. 
  We present the architecture of the proposed system and explain the working principle 
  of its constituent modules. Finally, we demonstrate the effectiveness of the proposed system through an evaluation experiment and a questionnaire.
\end{abstract}




\maketitle

\section{Introduction}
The consumption pattern of the consumers throughout the world has undergone a complete shift due to advances made in science and technology. 
The internet, for example, has pushed consumers away from physical stores and given them the freedom via e-commerce to shop anytime, 
from anywhere around the world, and without the need to leave their homes. Although this emerging scenario is advantageous to consumers in 
many aspects, for example, e-commerce ensures that consumers today have more choice than before, it also comes with some challenges. 
For example, it can become difficult for the consumers to find a high-quality product from the range of choices they have. Thus it is necessary to develop 
information retrieval systems for optimizing shopping experience.

According to our survey of the general shopping experience, it is very important to perform a comprehensive survey of the products to be purchased. The product 
information on the e-commerce website is relatively insufficient, and may not be able to fully explain the actual state of 
the products, and thus, it may become difficult for consumers to completely trust the e-commerce website. For example, 
consumers may be doubtful regarding the veracity of the product reviews, or they may have concerns about the quality of the product they may want to purchase. 
To address these challenges, this study develops an information retrieval system that can effectively improve the shopping retrieval experience 
by collecting information about products from multiple perspectives on the Internet.

The rest of this paper is organized as follows: Section 2 introduces the related works on content-based filtering and 
decision-making concept and provide an overview of the analytic hierarchy process (AHP) calculation process. Section 3 describes the system components 
in detail and explains its working mechanism. Section 4 evaluates the system by designing an experiment for evaluating the recommendation result 
generated by the system. Finally, we give a discussion in Section 5 and present our conclusions in Section 6.

\section{Related Work}

\subsection{Content-based Filtering}

Content-based filtering recommendation is a method used to search for similar items and recommend 
them to users by analyzing discrete features of the items centered on the user preferences. User preferences 
in this case are often informed in advance, and thus unlike collaborative filtering it does not suffer from the cold start problem. 
Thanks to the significant advancements made by the information retrieval and filtering communities and 
various text-based applications, many content-based systems focus on recommending items 
containing textual information, such as documents, websites (URLs), and Usenet news messages \cite{1423975}. 

In this study, we use the idea of content-based filtering to describe a product with textual information obtained from 
e-commerce websites and other multiple perspective sources.

\subsection{Multiple-criteria Decision-Making}

MCDM is a concept in decision science that is often used to solve multiple criteria 
decision or planning problems. Usually, there is no optimal solution for such problems, and it requires different solution 
routes to be selected according to the decision-maker's (DM's) preferences. Depending on whether the decision solution is 
finite or infinite, it can be broadly classified into two categories: multiple attribute decision making (MADM) and multiple 
objective decision making (MODM) \cite{zavadskas14}. Choosing the most suitable product is a very challenging task. 
It requires consumers to make comparisons among similar products and consider various different and possibly opposing principles, which 
makes it advantageous to use a multi-criteria decision-making strategy to select products.

The AHP, developed by Thomas L. Saaty, is a multiple-criteria decision analysis method used to analyze and organize complex decisions \cite{saaty1980analytic}. 
AHP is often applied to multi-objective, multi-criteria, multi-factor, multi-level unstructured and complex decision problems, especially strategic decision problem. 
Its flexibility and the fact that it can be easily combined with other methods make it extremely practical. AHP consists of three main operations \cite{saaty1980analytic}\cite{saaty2008decision}:
\begin{itemize}
  \item Hierarchy construction: This involves constructing a decision hierarchy, with the decision at the top, followed by the set of criteria in the intermediate level and the set of alternatives at the bottom level.
  \item Priority analysis: This involves constructing a set of pairwise comparison matrices, and using the priorities obtained from the comparisons to weigh the priorities in the level below.
  \item Consistency analysis: The consistency and coherence of the judgment are checked using the coherence ratio (CR). It is assumed that a CR equal to 
  0.10 or less is considered acceptable.
\end{itemize}

\section{Product Information Browsing Support System}

In this study, we will realize a system to support multifaceted information collection and product recommendation. The system will 
first filter unrelated products using an API based on a reference product specified by the user. The remaining products are called 
related products (RPs). Whether products are related to each other depends on whether they are semantically or physically 
related. For example, when the user selects "Google pixel 3" as the reference product, the system may search for the same brand of 
"Google pixel 4a", or it may search for other brands of cell phones "Apple iPhone 11", or it may search for an artwork "Pixel Art Screen", 
which has the keyword "pixel" but in reality is not a cell phone. These products are categorized as RPs. Then, the system 
searches for information around any RP in a multifaceted way. Here, "multifaceted" means that the system will not only
search information from e-commerce website, but also from several other channels that may contain information about the product 
(e.g., SNS website, online video platform, etc.). The information obtained by the system refers to a series of 
discrete information that can be used to describe the product, including the product title, rating, price on the e-commerce website, 
review counts, and so on. Finally, the system will evaluate these RPs based on user preferences. As it is difficult for a generic researcher 
to obtain large-scale data about user behavior, we cannot make recommendations based on user behaviors the way collaborative filtering 
algorithms do. Thus, in this study we use the techniques from decision science to address the issue of user preferences. Specifically, we use 
AHP to generalize attribute weights for each evaluation points and make a final score for each RP. The top n-ranked products are then recommended to users.

\subsection{System Architecture}

The architecture of the proposed system is shown in {\bfseries Figure 1}. This system is implemented in the form of browser extensions. The system 
is divided into three parts: the information preparation mechanism (marked in blue), the evaluation mechanism (marked in green), and the result 
presentation section (marked in orange). The information preparation mechanism comprises the product information preparation mechanism 
and the product video preparation mechanism. It retrieves product information for calculating similarity according to user input, checks 
whether there is a related video on the video site, and stores information about the video. 
\begin{figure}[t]
  \centering
  \includegraphics[width=\linewidth]{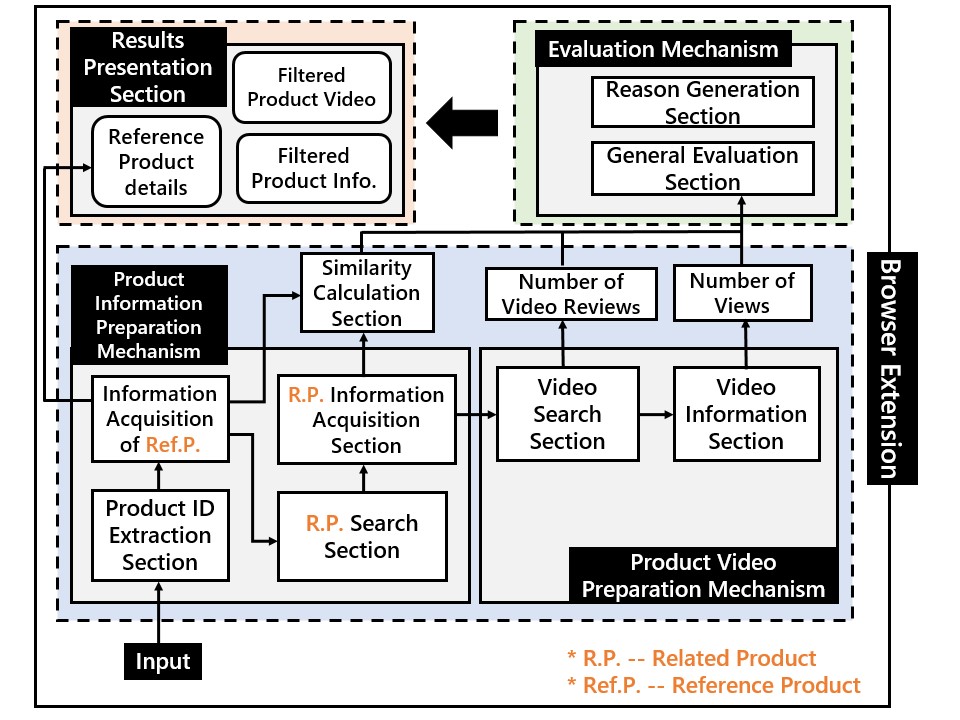}
  \caption{System architecture}
\end{figure}
The evaluation mechanism collects factors that affect the search results, and calculates the overall score of the 
recommended product by adjusting the weight of each factor according to the preferences of the user. It also uses the above information to 
generate the reason the product was displayed, and passes the information about the search product to the search product display mechanism. 
The search product display mechanism uses the necessary information to display the information in the browser extension.

\subsection{Similarity Calculation Module}

The similarity calculation module uses the product information obtained from e-commerce websites to calculate the similarity between the reference 
product and the RPs. As the variety of products on e-commerce websites is huge, we consider that it is too complicated to choose different attributes 
for different kinds of products to calculate the similarity. Thus, to ensure the applicability of the system, we decide to calculate the 
similarity between products based on two attributes: title and price.

In this study, we use cosine similarity to describe the degree of similarity between products. We assume that attribute vector $V_A$ of reference 
product $A$ is $(1, 1)$ and attribute vector $V_B$ of related product $B$ is $(v_t, v_1)$. The attribute vector of a product is 
a 2-dimensional vector space, where $v_t$ refers to the title vector and $v_1$ refers to the price vector. The formula for $v_1$ is shown below:
\begin{equation}
  v_1 = 1 - \frac{\left\lvert P_{B_k} -P_A \right\rvert }{range(P)}
\end{equation}
where, $B_k$ is an element of RP set $\mathcal{B} = \{B_1, B_2, ..., B_n\} $, $B_k \in \mathcal{B}$. $P_X$ refers to the price of product 
$X$. $range(P)$ refers to the distance between the maximum dense intervals of the mathematical distribution of prices.

We use the cosine similarity between reference product $A$'s word vector $W_A$ and related product $B_k$'s word vector $W_{B_k}$ to represent $B_k$'s 
title vector $v_t$. According to the word vector $W_X$ of any product $X$, we use term frequency inverse document frequency ({\bfseries TF-IDF}) to calculate the 
importance of each word in the title of $X$. The vector formed by enumerating the importance values of each word is $W_X$. For example, the title of product 
$X$ is "Best soft drink in Japan", where the importance of each word are "0.56", "0.78", "0.61", "0.32" and, "0.45", respectively, and thus word vector $W_X$ will be defined 
as [0.56, 0.78, 0.61, 0.32, 0.45]. {\bfseries TF-IDF} is a numerical statistic that is intended to reflect how important a word is to a document in a 
corpus. It is calculated as:
\begin{equation}
  tfidf(t,d,D) = tf(t,d) \cdot idf(t,D)
\end{equation}
where
\begin{itemize}
  \item $tf(t,d)$ is the term $t$'s frequency adjusted for document $d$'s length (in this study, it refers to the frequency of each word in title)
  \item $idf(t,D)$ is a measure of how much information the word provides ( obtained by dividing the total number of documents containing the term, and then 
  taking the logarithm of that quotient).
\end{itemize}
After obtaining both $W_A$ and $W_{B_k}$, we can obtain $v_t$ by calculating $cos(W_A, W_{B_k})$ as shown below:
\begin{equation}
  similarity = cos(A, B) = \frac{\sum_{i = 1}^{n} A_iB_i }{\sqrt{\sum_{i = 1}^{n} A_i^2 } \sqrt{\sum_{i = 1}^{n} B_i^2}}
\end{equation}
So far, we have obtained a complete attribute vector $V_B = (v_t, v_1)$. We, next, calculate the cosine similarity between reference product $A$ and RP $B$ to obtain the similarity between them.

\subsection{Evaluation Module}

The module is divided comprises: a comprehensive evaluation section (CES) and a result interpretation section (RIS). The CES evaluates each RP from multiple 
perspectives, generates corresponding weights for each indicator according to user's own preferences, and finally gives a comprehensive score for the product 
according to our proposed rating criteria. The RIS explains why the product is recommended by presenting the rating data obtained from the CES to the user in the 
form of a graph.

{\bfseries Comprehensive evaluation section (CES)}. Our system makes general recommendation of all existing products on the e-commerce website. For the CES, 
instead of selecting evaluation criteria for a specific category of goods, we select a set of evaluation criteria that are universally applicable to all goods. Thus, we choose 
five criteria: similarity (SI), number of reviews (NR), rating (RA), number of video reviews (NVR), and number of video plays (NVP). Here, SI is 
calculated as shown in {\bfseries 3.2}, which narrows the range of RPs. NR and RA are two very common and valid evaluation criteria on e-commerce websites, while NVR and NVP are the number of reviews 
and plays corresponding to the most popular videos that can be searched on video sites, which represent objective evaluation criteria from other information sources. The alternatives is a group 
of n RPs, where n fluctuates between 20 and 30 per search, depending on product category. The proposed AHP model is as shown in the {\bfseries Figure 2}.
\begin{figure}[t]
  \centering
  \includegraphics[width=\linewidth]{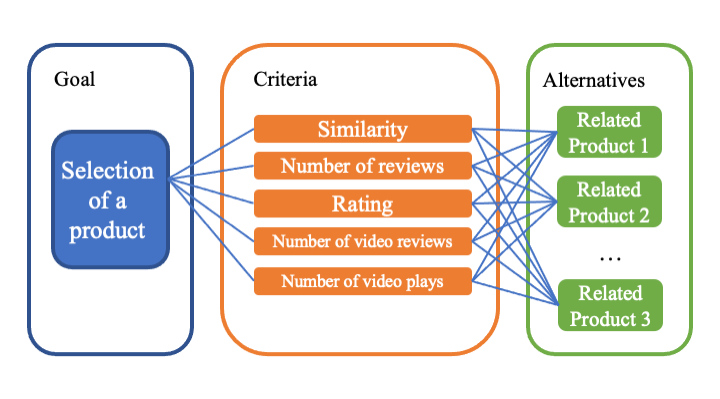}
  \caption{The proposed AHP model for choosing among several related products}
\end{figure}

Based on our design and understanding of what consumers focus on when shopping, we suggest as an illustration the pairwise comparison of the criteria. An example is as shown in {\bfseries Table 1}. 
As it is proposed in the paper of AHP \cite{saaty1980analytic}, the $\lambda_{max}$ is called principle Eigen value, which is obtained from the summation of products between each element of Eigen vector and the 
sum of columns of the reciprocal matrix. The $CI$ is a measure of consistency, called Consistency index as deviation, which is calculated using $\lambda_{max}$. The $CR$ is called Consistency Ratio, which is a comparison 
between Consistency Index and Random Consistency Index. If the value of Consistency Ratio is smaller or equal to 0.1, the inconsistency is acceptable. If the Consistency Ratio is greater 
than 0.1, we need to revise the sujective judgment (the pairwise comparison matrix). Here the $CR$ is less than 0.1, which means the result is acceptable. {\bfseries Table 2} shows the weights of the criteria. 
For this example, the criteria that the user is most interested in are NR, SI and NVR.
\begin{table}
  \caption{Pairwise comparison of the criteria}
  \label{tab:freq}
  \begin{tabular}{l*{5}{c}}
    \toprule
     &SI&NR&RA&NVR&NVP\\
    \midrule
    SI &1 & 1/3 & 7 & 3 & 5\\
    NR &3 & 1 & 9 & 5 & 7\\
    RA &1/7 & 1/9 & 1 & 1/5 & 1/3\\
    NVR &1/3 & 1/5 & 5 & 1 & 3\\
    NVP &1/5 & 1/7 & 3 & 1/3 & 1\\
  \bottomrule
  $\lambda_{max} = 5.2372$ & $CI = 0.0593$ \\ $CR = 0.0529$
\end{tabular}
\end{table}

\begin{table}
  \caption{Weights of the criteria}
  \label{tab:freq}
  \begin{tabular}{l*{2}{c}}
    \toprule
     & Criteria & Weight\\
    \midrule
    1 & SI & 0.2638\\
    2 & NR & 0.5100\\
    3 & RA & 0.0329\\
    4 & NVR & 0.1295\\
    5 & NVP & 0.0636\\
  \bottomrule
\end{tabular}
\end{table}
After obtaining the weights of each criteria, we introduce the scoring principles for each criteria to calculate a comprehensive score of each product. For both SI and RA, 
we simply convert them to percentages to obtain the scores. For example, for a RP with a similarity of 0.82 and a rating of 4.7/5.0, the scores of these two criteria will be 82 and 94, respectively. 
For NR, NVR and NVP, first we need to set thresholds for them based on the actual data sources. Then we multiply the value of the criteria as a percentage of the overall evaluation interval by 100 
as the score for that criteria. For example, as the source of NVP in our experiment is YouTube plays, we think that over $10^5$ plays can be considered as very popular which is 100 points. Thus, 
the evaluation interval of NVP in our experiment is $(0,10^5)$, if we have a video of 6,3850 plays, the score of NVP for this video will be $score(5) = 63850 / 10^5 \cdot 100 = 63.85$. 

\begin{equation}
  Score = \sum_{i = 1}^{n}  weight(i) \cdot score(i)
\end{equation}

After obtaining the scores for each criterion, we can now use the weights shown in {\bfseries Table 2} to calculate the comprehensive score. The formula is shown as (4).

{\bfseries Result interpretation section (RIS)}. Based on the scores of each criteria obtained by the CES and the final comprehensive score, we present the results to the user in the form of progress bars. 
We believe this helps the user to understand the reason why the search result is being presented to him/her. By comparing the difference between each branch and the overall score, users can intuitively 
access the strengths and weaknesses of the product and make a reasonable judgment.

\subsection{User Interface}

The installation of the proposed system is shown in {\bfseries Figure 3}.
\begin{figure}[tbp]
  \centering
  \includegraphics[width=\linewidth]{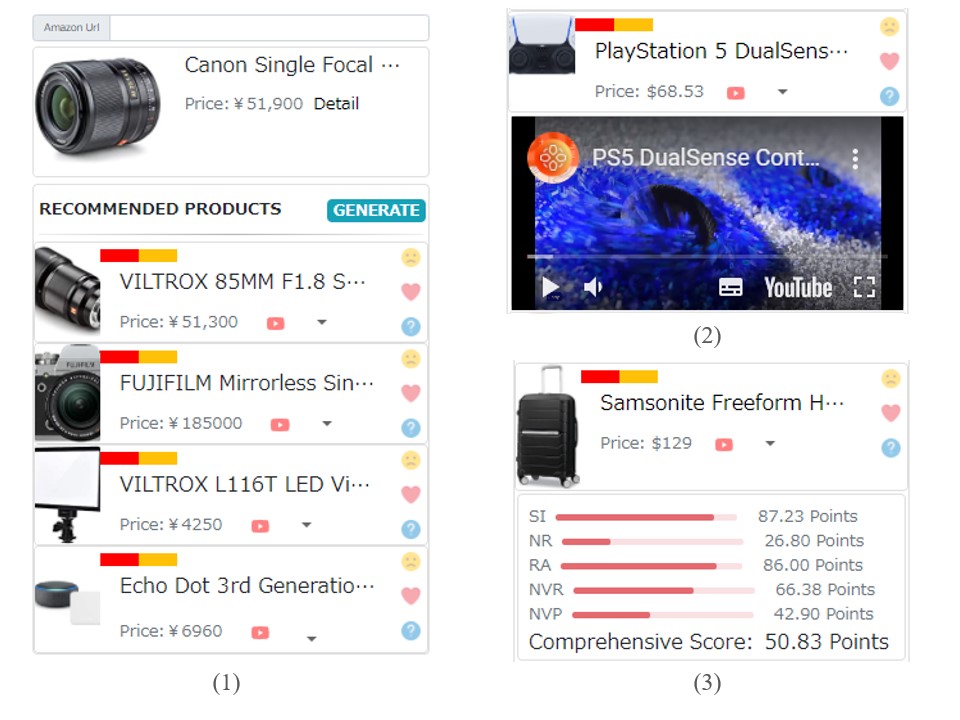}
  \caption{User interface of the proposed system}
\end{figure}
Part (1) shows the initial interface of the system, with an input box at the top to enter a link to a reference product, which is then recognized by the system in real time, and its 
information is displayed below. The bottom half of the interface is used to display the recommended products, and the green generation button in the middle of the interface 
controls the generation of recommendation results. The recommended products are presented on the interface in cards, each with basic product information, video button, detail button, 
and favorite, hate, result interpretation buttons on the sidebar. The response result of the video button is shown in part (2), and the presentation of result interpretation is as 
shown in part (3).

\subsection{Installation Details}
About technical details, we choose Google's Chrome browser to develop our browser extension\footnote{https://developer.chrome.com/docs/extensions/}. 
We use a fee-based API "Amazon Product / Reviews / Keywords API" on RapidApi website \footnote{https://rapidapi.com/?site} to get the product data from Amazon. The API service has four 
data interfaces, namely, product search, product details, product reviews, and categories acquisition. Specifically, in this system, we first extract the product asin number (a unique 
identifier for each product) from the reference product link entered by the user and input it as a parameter to the product details API to get the details of the reference product. 
Based on the product's category obtained through categories acquisition API, we use the product searching API to search for related products of the reference product. We use Google's YouTube 
Data API \footnote{https://developers.google.com/youtube/v3} to search for videos of related products on the YouTube video site, and select the video with the highest number of plays as 
the representative video of the product by adjusting the parameters of the API, and record the criteria value of the video. Finally, we use all the obtained data to conduct a comprehensive 
analysis by using the methods in {\bfseries 3.2} and {\bfseries 3.3}. As for the visual design of the system, we use a free and open-source CSS framework called Bootstrap \footnote{https://getbootstrap.com/} to realize the user interface as it is shown in {\bfseries 3.4}.

\section{Evaluation}

\subsection{Experiment of Result Evaluation}
To evaluate the results generated by this system, we investigate the levels of user satisfaction with the results in real situations through an experiment and a questionnaire. 
For this purpose, we designed multiple sets of controlled trials. Specifically, we selected one representative product from four domains: apparel (AP), electrical appliances (EA), furniture (FU), 
and food (FO) as the reference product, and generated the search results through three different methods. For each respondent, we shuffled the results for each domain and asked subjects to rank the results 
according to their satisfaction without telling them in advance the method used for each result. The process of the experiment is shown in {\bfseries Figure 4}, 
where white, grey, and black cubes represent the generators using three different methods, respectively. The white method searches only
based on the similarity calculated by title vector $v_t$ and price vector $v_1$ of the product. The grey method searches with equal weights based on information from multiple perspectives 
such as similarity, number of reviews, etc. Finally, the black method searches using the weights calculated by AHP based on information from multiple perspectives. Each 
respondent will be asked to rank the three groups of results for each of the four domains. AHP should be applied to each respondent before experiment, so that we can obtain his/her comparison matrix for applying our method.
\begin{figure*}[htbp]
  \centering
  \includegraphics[width=\linewidth]{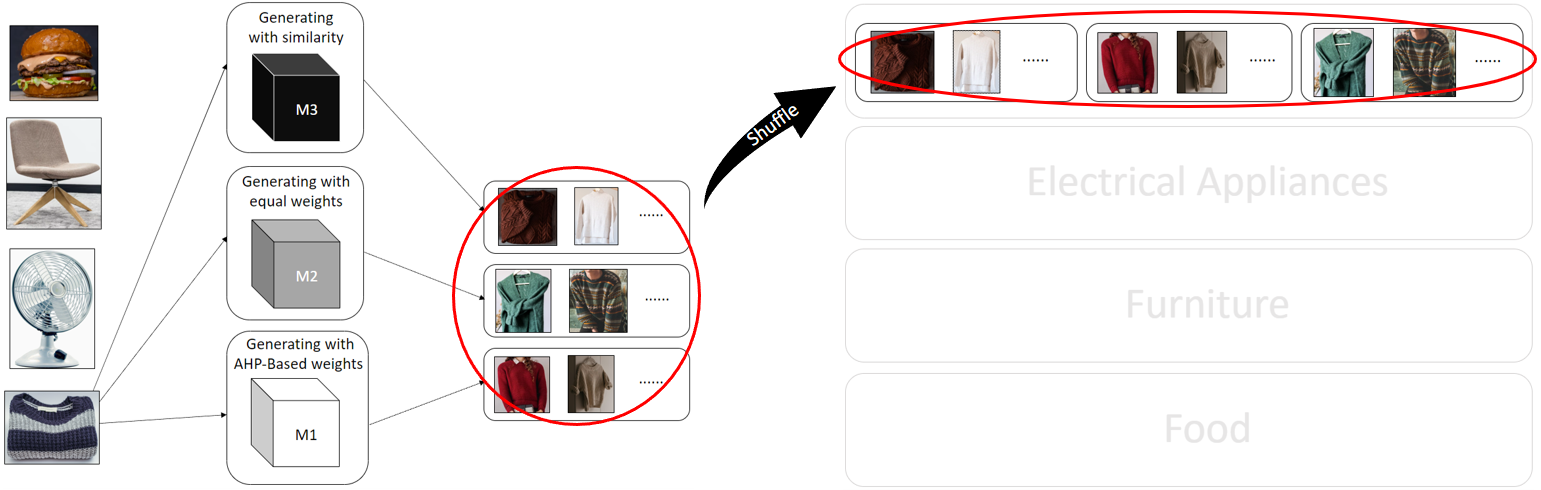}
  \caption{The process of the experiment}
\end{figure*}

At the same time, to have a deeper understanding of the respondents' intuition about the search results of this system, we want them to answer several system-related questions 
after completing the above experiment. We referred to the design ideas of previous studies \cite{pu2011user, Silveira2019HowGY, knijnenburg2012explaining} and design a questionnaire suitable for our system. respondents 
are asked to indicate their answers to each of the questions using a 1-5 Likert scales, where 1 represents "strongly disagree" and 5 represents "strongly agree". The items of the questionnaire are shown below:
\begin{enumerate}
  \item The interface provides sufficient information.
  \item I know exactly what kind of items I like.
  \item I can clearly feel the differences between three results of each domain.
  \item My selection criteria for shopping will be very different at different times.
  \item I'm used to collecting all kinds of information when buying items and selecting carefully.
  \item I trust my own findings more than I trust the machine's recommendations.
\end{enumerate}

We selected 20 respondents between the ages of 20 and 30, including 6 women and 14 men. The result of the first sorting experiment is shown in {\bfseries Table 3}. 
It can be seen that for any of the four domains, according to the proportion of the results in which respondents perceive the highest level of satisfaction, our system is the highest. 
In other words, the results obtained from a multiple perspective search are more likely to be favored by users than those from a method that considers only similarity. Meanwhile, the results 
obtained by considering users' decision propensity cannot be ignored. It is worth noticing that the satisfaction level obtained with our system in the FO domain is not very high, although 
it is higher compared with the other two methods. We asked some respondents afterwards, and the possible reasons for the results are that everyone has different natural food preferences, and 
that it's hard to judge by packaging alone without tasting it yourself.
\begin{table}
  \caption{The percentage of different satisfaction levels achieved by the proposed system in each case (AP: apparel, EA: electrical appliances, FU: furniture, FO: food)}
  \label{tab:freq}
  \begin{tabular}{l*{4}{c}}
    \toprule
     & Highest & Middle & Lowest\\
    \midrule
    AP & 70\% & 15\% & 15\% \\
    EA & 75\% & 15\% & 10\% \\
    FU & 80\% & 15\% & 5\% \\
    FO & 55\% & 30\% & 15\% \\
    Total & 70.00\% & 18.75\% & 11.25\% \\
  \bottomrule
\end{tabular}
\end{table}

The result of our questionnaire is shown in {\bfseries Figure 5}. We use bar charts to describe the overall distribution of the sample. In Q1, we asked respondents whether the information 
cues in this experiment were sufficient enough for them to make a decision. More than half of them gave a positive answer, some gave a neutral option, and only 
a few raised an objection. On the whole, this can be considered a relatively fair experiment. For Q2, two groups with similar numbers of people chose opposite attitudes. 
Similar to most shopping scenarios, not everyone knows exactly what kind of products they want most, but we can determine whether the search results meet our expectations 
based on a certain set of information, and consistently receiving meaningful information can improve their user experience. Therefore, for Q3, most of the respondents think they can find or feel the 
difference between the three results through intuition. The results of the first three questions show that users have the ability to discriminate between good and bad results. Q4, Q5, and Q6 revolve 
around the value-in-use aspect of the system. The answers for Q4 and Q5 indicate that there exist some people whose selection criteria for shopping will change over time, and that 
most of them are used to wasting their time investigating about the quality of their buying target. Our system can save most of the information gathering work 
for users and can make customized recommendations based on different people's decision-making habits. From the responses to Q6, it was clear that the consumers' trust in machine-based 
recommendations is increasing in parallel to the advancements made in technology.
\begin{figure}[t]
  \centering
  \includegraphics[width=\linewidth]{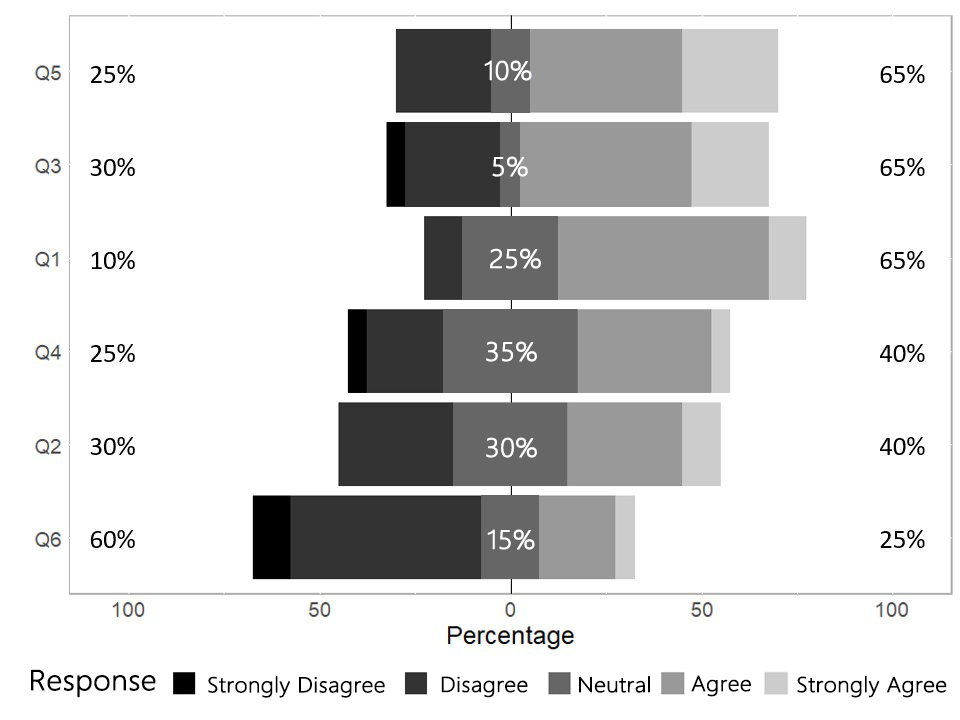}
  \caption{Statistical results of the questionnaire}
\end{figure}

\subsection{System Response Time}

Considering the user experience in real-world scenarios, we also measured the response time of each display function of the system. There are two aspects of the proposed system that may turn out 
to be somewhat time-consuming. First is the time it takes to enter the reference information until the reference product information is fully displayed. The second is the time it takes for the system to 
display all the results after clicking the "Generate" button. We named these two parts as "Reference Product Information Display Time (RPIDT)" and "Results Generalization Time (RGT)", and tested each 50 times. 
The mean time of RPIDT is 2843.3 ms and that of RGT is 3127.5 ms. The results indicate that the response time for both these processes is around 3s, which is within the tolerable range of users. The user experience of our system is qualified on the point of system response time.

\section{Discussion}
Usually, the buying decisions on e-commerce platforms are subject to different user preferences. In this study, we tried a different approach. We inform the consumers about each buying criteria and use AHP to help the calculate the weight of each criteria in their mind. However, it would 
be helpful to further investigate whether there exists a generally appropriate weighting that can satisfy all consumers in a certain domain.

The response time of each function of our system is around 2 to 3s, which mainly depends on the API used to request data. For web services, a comfortable 
response time should be in milliseconds, from this perspective, our system does not perform well and should be combined with UI 
optimization to improve user experience. However, from the usage results perspective, the total waiting time of the proposed system is approximately 6s. If users search for their favorite products on their 
own without the help of the system, it often takes more than 10 minutes or even longer, and thus, the proposed system is successful in 
helping users cut down their searching time.

In this paper, we do not give any comparative evaluation experiment with other approaches. This is because most of the mainstream recommendation approaches are implemented based on 
traditional models or deep learning models with ideas such as collaborative filtering, matrix factorization, logistic regression, etc. MCDM analysis is a sub-discipline of operation research and the domain 
itself does not overlap with them, thus comparisons cannot be made. At the same time, AHP is a subjective decision making approach compared to other MCDM approaches, so it is also difficult to perform comparative 
analysis with other MCDM approaches \cite{jtaer16060122}.

\section{Conclusion}

We draw on the idea of content-based filtering and apply a the MCDM technique to implement a product information retrieval system. We developed a browsing support system for obtaining product information 
from multiple perspectives and obtain user preference weights based on AHP techniques to comprehensively evaluate each related product and finally display the top-rated products to users. Through 
evaluation experiments, we verified the effectiveness of the system in improving the quality of retrieval results. However, for the shopping process itself, there still are areas that need further consideration.

\begin{acks}
This work was supported in part by JSPS KAKENHI Grant Number JP19K12266.
\end{acks}

\bibliographystyle{ACM-Reference-Format}
\bibliography{sample-base}

\end{document}